# Observation of transferred-electron oscillations in diamond

N. Suntornwipat[1], S. Majdi[1], M. Gabrysch[1], I. Friel[2] and J. Isberg[1]

[1]*Division for Electricity, Department of Engineering Sciences, Uppsala University, Box 534, 751 21, Uppsala, SWEDEN*
[2]*Element Six Innovation, Fermi Avenue, Harwell Oxford, Didcot, Oxfordshire OX11 0QR, UK*
(Dated: 26th Aug 2019)

The transferred-electron oscillator (TEO) is a device used in microwave applications that utilizes the negative differential mobility (NDM) effect to generate continuous oscillations. Recently, NDM was observed in intrinsic single-crystalline chemical vapor deposition (SC-CVD) diamond. The occurrence was explained by the electron repopulation between its different conduction band valleys. This paper presents the results of constructing a diamond TEO based on the NDM effect. A series of experiments has been performed for varying voltages, temperatures and resonator parameters on three SC-CVD diamond samples of different thicknesses. For the temperature range 90 – 300 K, we observe transferred-electron oscillations in diamond.

*Introduction* – Transferred-electron devices (TEDs) or Gunn diodes are widely used in microwave applications as local oscillators and drivers for amplifier chains. [1,2] The TED is one type of negative resistance oscillator which uses negative differential mobility (NDM), caused when electrons are scattered from the central conduction band valley with low effective electron mass to satellite valleys with high effective electron mass, to generate continuous oscillations. Oscillations in InP and GaAs devices were first reported by J.B. Gunn [3,4] in 1963, but he did not give an explanation for the phenomenon at the time. However, it turned out that oscillations had already been predicted by C. Hilsum [5] in the previous year, based on a suggestion by B.K. Ridley and T.B. Watkins [6] that certain materials should exhibit the NDM effect. Gunn oscillations are usually only observed in direct bandgap materials such as GaAs, InP, InAs, CdTe and ZnSe [3,7–9] and is not normally associated with elemental or indirect-bandgap semiconductors.

Diamond exhibits many exceptional material properties such as high thermal conductivity, extreme mechanical strength, high carrier mobilities and chemical inertness. Together with the possibility to synthesize high-purity single-crystalline chemical vapor deposition (SC-CVD) diamond [10], it is a very interesting material for semiconductor devices. Since diamond is an indirect semiconductor without a central (Γ-point) valley in the conduction band, one might not expect that this material exhibits the NDM effect. Nevertheless, the first observation of NDM occurring in intrinsic SC-CVD diamond was reported in ref. [11] at a temperature range of 110 to 140 K for an applied electric field of 300 to 600 V/cm in the [100] direction.

Even though diamond has energetically equivalent valleys, NDM occurs because of the anisotropic transport properties of electrons in different valleys, i.e. by different electron effective masses in different directions. The six conduction band valleys in diamond are situated at 76% of the distance from the Γ-point on the Δ-axis in the Brillouin zone (BZ), as illustrated in Figure 1a. [12–14] When applying an electric field in one direction (in the present case along [100]), electrons in the two valleys aligned in parallel (on the (100) axis) to the field and the four valleys aligned orthogonally (on the (010) and (001) axes) to the field respond with different effective masses. The longitudinal electron effective mass ($m_l^*$) is $1.56m_0$ and the transverse electron effective mass ($m_t^*$) is $0.28m_0$. [15] In contrast, in the case of GaAs, NDM is explained by the repopulation of electrons between a central conduction band valley (Γ-point) and satellite valleys (L-point), with different energy minima (about 300 meV). [16] This is shown in Figure 1b.

The possibility of using the NDM effect in diamond devices has been investigated with a theoretical model to establish if it is feasible to demonstrate a diamond based transferred-electron oscillator (TEO). [17] In this model the TED operates with an electric field applied in such a way that the internal space charge distribution and the internal electrical field distribution become unstable and cause oscillations in a resonant circuit. This paper presents experimental results of an actual diamond-based TEO, where oscillations have been observed for the temperature range 90 to 300 K. This is the first time oscillations have been observed in a bulk TEO based on an elemental semiconductor or an indirect-bandgap semiconductor.

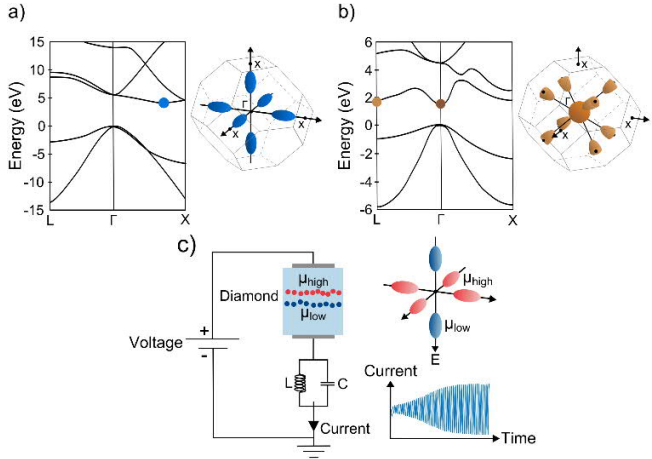

Figure. 1. The band structure of (a) diamond, and for comparison (b) GaAs. The valleys in diamond are located along {100} directions near the zone boundary and have the same minimum energy, 5.47 eV above the valence band maximum. The 3D pictures show equal-energy surfaces in the first Brillouin zone surrounding the conduction band minima. When applying an electric field along the [100] direction, the two valleys aligned in parallel to the field and the four valleys aligned orthogonal to the field respond with different electron effective masses and mobilities. In GaAs, the central valley ($\Gamma$) lies about 300 meV below the ellipsoidal satellite valleys (L) situated along {111} directions. Gunn oscillations in GaAs instead arise from electron repopulation between central and satellite valleys. (c) A diamond Gunn oscillator where the oscillations in current are the results of electrons repeatedly depleting and accumulating inside an intrinsic diamond film.

*Experiment* – The schematic of the measurement setup is shown in Figure 2. A HeAg excimer laser with a wavelength of 224.3 nm and a peak power of approximately 50 mW is used to generate electron-hole pairs. Photons with this wavelength are absorbed according to the Lambert-Beer law with a penetration depth of around 200 µm in intrinsic diamond. [18]

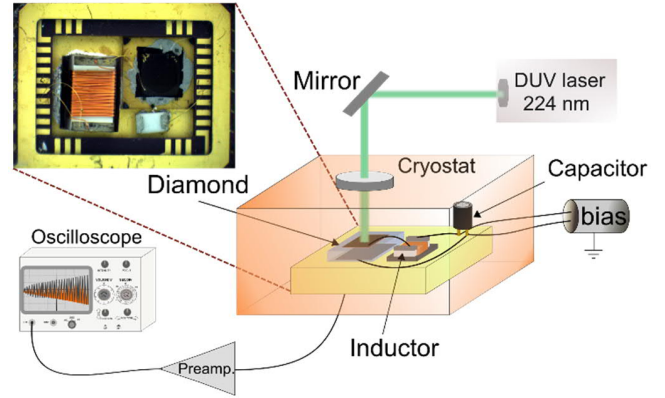

Figure. 2. Simplified schematic of the measurement setup used for testing the TEO devices. The inset shows a diamond sample mounted in a chip carrier together with a resonant LC-circuit.

A pulse length of 100 µs is used, which is much longer than the electron transient time, which occurs on a 10 to 100 ns timescale in the experiments. A bias voltage is applied between the front (the illumination side) and the back contact, with the negative bias at the front in order to make the electrons travel through the sample and to extract the holes at the illumination side. The circular contacts are 3 mm in diameter and are, following the standard optical lithography techniques, deposited by sputtering with Ti/Al (20/300 nm) to form a semitransparent mesh on the (100) surface. This geometry makes it possible to achieve a nearly homogeneous electric field and also permits UV photons to enter the region underneath the contact. The sample is mounted on a gold plated chip sample holder where it is connected in series with an LC circuit consisting of a ferrite core inductor (L) and a capacitor with NP0 dielectric (C), (see Figure 1c). The sample holder is mounted on the cold finger of a vacuum cryostat. The current through the sample and LC circuit is amplified, using either a commercial amplifier placed outside the cryostat (Mini-Circuits ZFL1000LN+), or a custom-built p-HEMT based amplifier cooled to sample temperature inside the cryostat. The amplified current is recorded with a digital oscilloscope (Tektronics TDS 684C, 5 GS/s). A doubly shielded casing is used to carefully isolate the sample and the HEMT amplifier from any radio frequency noise generated in the outside environment. To stabilize the bias voltage a capacitor of 33 nF with NP0 dielectric is added in close proximity to the sample. A calibrated GaAlAs diode sensor (TG-120-CU-HT-1.4H) in contact with the sample holder is used for temperature monitoring in conjunction with a Lake Shore 331 temperature controller. Liquid nitrogen is used as the coolant.

*Results and discussion* – At low temperatures, intervalley scattering caused by short wavelength lattice vibrations are absent in diamond due to the extreme rigidity of the lattice. This makes it possible to observe electrons moving with different velocities depending on their valley state. At low electric fields, all valleys are roughly equally populated and the electron drift velocity is proportional to the field. As the applied electric field is increased, the electrons in orthogonal valleys are heated up more than the electrons in parallel valleys due to the difference in their effective masses along the direction of the field. The electrons from the orthogonal valleys reach the energy threshold for Longitudinal Acoustic (LA) phonon emission of 120 meV first, which leads to intervalley scattering and a strong repopulation of conduction band electrons from orthogonal to parallel valleys. This phenomenon causes the observed negative differential mobility. As the electric field is increased further, electrons in the parallel valleys will reach a sufficient energy to re-scatter into the orthogonal valleys, which tends to equalize the valley population. If NDM is present and the carrier concentration exceeds a certain threshold, instabilities in the spatial carrier distribution can occur as fast electrons catch up with slow electrons. This causes a build-up of an accumulation layer and at the same time leaves a depleted region behind. The internal space charge distribution of such accumulation and depletion of electrons results in an oscillating current. [19] Figure 1c shows a Gunn oscillator where the oscillations in current are the results of electrons repeatedly depleting and accumulating inside intrinsic diamond.

Experiments were performed at a temperature range of 90 to 300 K using three high-purity SC-CVD diamond samples from Element Six Ltd. with sample thicknesses of 307, 390 and 420 µm. The details of the samples used for this study are summarized in table 1. The samples were illuminated by the laser and the induced current was measured. Several different experiments were performed with inductance and capacitance values of external LC circuit varying between 1.8 and 10 µH and 0 and 4.7 pF, respectively. For C = 0 the resonance relies on the internal capacitance of the sample (approx. 0.5 pF) and stray capacitance in the sample holder. For the temperature range where NDM had previously been reported we observed oscillations in the frequency range of 20 to 50 MHz depending on the chosen inductor and capacitor values. Figure 3 to Figure 5 show the measured current from three such tests. Here, the 307 and 390 µm samples were separately connected in series with an inductor of 5.6 µH (devices A and B respectively), while the 420 µm sample was connected in series with an inductor of 10 µH (device C). No external capacitor was connected in either case and bias voltages of 0 to 30 V were applied across the devices.

TABLE I. Details of the diamond samples studied in this work.

| Sample | Description | Size [mm] | Thickness [µm] | Impurity Conc.[cm$^{-3}$] |
|---|---|---|---|---|
| #A | SC-Intrinsic | 2.8 x 2.8 | 307 | <10$^{14}$ |
| #B | SC-Intrinsic | 3.5 x 3.5 | 390 | <10$^{14}$ |
| #C | SC-Intrinsic | 4.5 x 4.5 | 420 | <10$^{14}$ |

The current waveforms at the temperatures 90 and 110 K with bias voltages of 0 and 20 V are shown for device A in Figure 3. With a voltage applied, current oscillations are observed having a much larger amplitude at 110 K compared to 90 K. A comparison of the current waveforms in the frequency domain at different voltages and at temperatures of 90, 110 and 290 K are presented for device A in Figure 4. The oscillations are strongest at 110 K and are substantially weaker at 90 or 290 K. By increasing the voltage, oscillations with larger amplitude are observed. Figure 5 shows a comparison of the Fourier transform of the current as a function of frequency and temperature for devices A, B and C at a bias voltage of 20 V. Device A, based on the thinnest sample, exhibits the strongest oscillations of the three devices. It should be noted that the oscillations are observed outside the NDM region (110 to 140 K), but rapidly diminish as the temperature is lowered below 100 K.

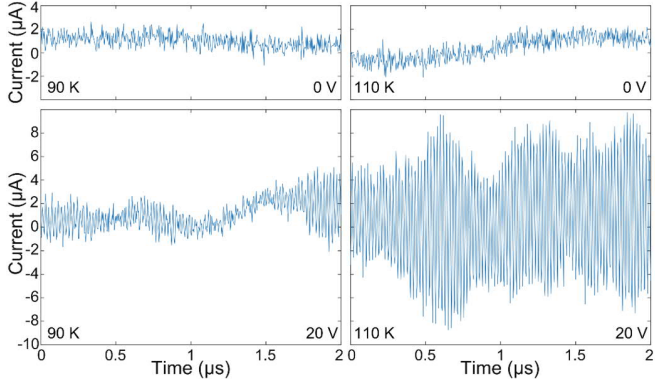

Figure. 3. The current waveform obtained at a temperature of 90 and 110 K and at a bias voltage of 0 and 20 V. The measurements were performed on device A.

In all three cases the oscillator strength is at its maximum at 110 to 130 K, as would be expected from simulations [17]. This can be explained by the negligible rate of intervalley phonon scattering at low temperatures which effectively inhibits valley repopulation. [12] Oscillations are however also observed well above 150 K and their amplitude only slowly diminishes with increasing temperature. Even at room-temperature weak oscillations can be observed. Clearly the model presented in Ref. [17] does not explain the high-temperature behavior. In this model, the population ratios are defined by first order rate equations, and we believe that this approach may be too simplistic to capture the detailed *dynamics* of the valley population. Presumably a more comprehensive model that includes a full description of the population dynamics and the scattering rates, possibly using ensemble Monte-Carlo simulations, is necessary to explain the originate of the oscillations observed at higher temperatures.

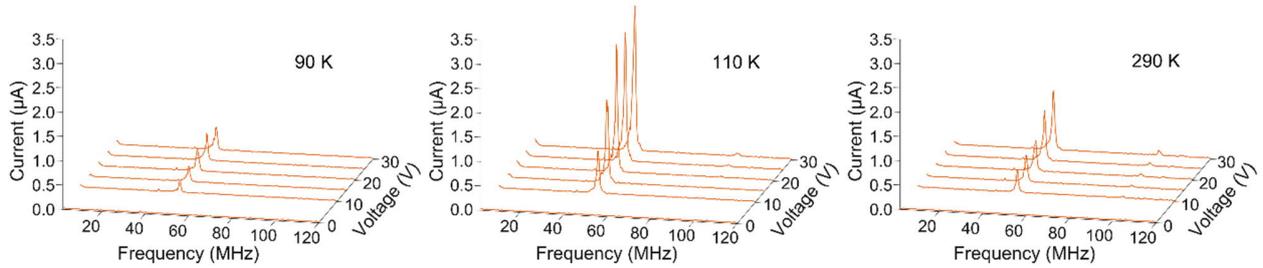

Figure. 4. The Fourier transform of the measured current as a function of frequency and bias voltage at a temperature of 90, 110 and 290 K. The bias voltage varied between 0 and 30 V. The measurements were performed on device A.

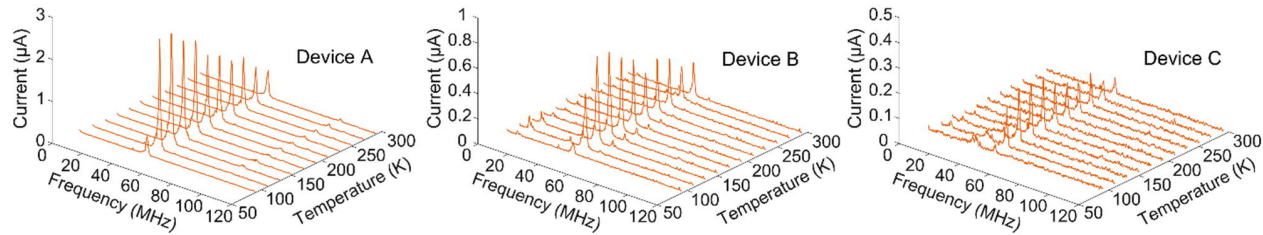

Figure. 5. The Fourier transform of the measured current as a function of frequency and temperature at a bias voltage of 20 V. The temperature varied between 90 and 300 K.

*Conclusion* – For a transferred-electron device based on approximately 400 µm thick diamond sample connected in series with an LC resonance circuit Gunn oscillations were observed for a temperature range of 90 to 300 K. For all our devices, the oscillations were most pronounced for temperatures between 110 and 130 K in the frequency range of 20 to 50 MHz, depending on the chosen inductor and capacitor values of the LC circuit. These oscillations can be explained by electron density instabilities occurring in situations with negative differential carrier mobility. Surprisingly, weak oscillations have been observed also at temperatures above 150 K and we suggest that a detailed model that includes a full description of the population dynamics and the scattering rates could give an explanation for this observation.

*Acknowledgement* – The authors would like to thank the Swedish Research Council (VR, Grant No. 621-2012-5819, 621-2014-6026 and 2018-04154) for financial support. In addition, Swedish Energy Agency (Grant No. 44718-1) is gratefully acknowledged by the authors. SM would like to thank Prof. S. Shikata (AIST) for providing the opportunity to conduct part of the research in his group and express his gratitude to the Japan Society for the Promotion of Science (JSPS) for financial support.